\documentstyle[12pt]{article}
\newcommand{\bec}{\begin{center}}
\newcommand{\ec}{\end{center}}
\newcommand{\bee}{\begin{equation}}
\newcommand{\ee}{\end{equation}}
\newfont{\blackboard}{msam10 scaled\magstep2}
\newcommand{\gf}{\mbox{\blackboard\symbol{'002}}}
\newcommand{\cf}{\mbox{\blackboard\symbol{'034}}}
\newcommand{\cg}{\mbox{\blackboard\symbol{'035}}}

\begin{document}
\large
\begin{titlepage}
\bec
{\Large\bf  D-Branes and Derived Categories \\}
\vspace*{15mm}
{\bf Yuri Malyuta \\}
\vspace*{10mm}
{\it Institute for Nuclear Research\\
National Academy of
Sciences of Ukraine\\
03022 Kiev, Ukraine\\}
e-mail: interdep@kinr.kiev.ua\\
\vspace*{35mm}
{\bf Abstract\\}
\ec
The digest of ideology interpreting 
D-branes on Calabi-Yau manifolds 
as objects of the 
derived category is given.
\vspace*{1cm}
\hspace*{2.3cm} Monodromy.
\end{titlepage}
\section{Introduction}
Recently there has been substantial 
progress \cite{1.,2.,3.} in 
understanding D-branes on Calabi-Yau
manifolds in context of derived categories 
\cite{4.}.

	The purpose of the present paper is
to give the digest of this ideology.
\section{Sheaves}
In this section we shall introduce the
definitions of presheaves and sheaves
\cite{5.}.

	A {\it presheaf} $F$ over
a topological space $X$ is 

	1)\ An assignment to each nonempty
open set $U\subset X$ of a set $F(U)$ 
({\it sections} of a presheaf $F$);

	2)\ A collection of mappings
(called restriction homomorphisms)
\[r_{UV}: F(U)\rightarrow F(V)\]
for each pair of open sets $U$
and $V$ such that $V\subset U$,
satisfying
\[r_{UU}=1_{U}\ ,\ \ \ r_{VW}\ r_{UV}=r_{UW}\ \ \
for\ \ W\subset V\subset U\ .\]

	A presheaf $F$ is called
a {\it sheaf}\ \ if for every
collection $U_{i}$ of open subsets
of $X$ with $U=\bigcup
\limits_{i\in I}U_{i}$ the following
axioms hold\ :

	a)\ If $s, t \in F(U)$ and
$r_{UU_{i}}(s)=r_{UU_{i}}(t)$ for
all $i$, then $s=t$\ ;

	b)\  If $s_{i}\in F(U_{i})$
and if for $U_{i}\bigcap U_{j}\neq \emptyset$
we have 
\[r_{U_{i},U_{i}\bigcap U_{j}}(s_{i})=
r_{U_{j},U_{i}\bigcap U_{j}}(s_{j})\]
for all $i$, then there exists an 
$s\in F(U)$ such that $r_{U,U_{i}}(s)=s_{i}$
for all $i$\ .

	If $F$ and $G$ are presheaves
over $X$, then a {\it morphism}
of presheaves $f: F\rightarrow G$
is a collection of maps
$f(U): F(U)\rightarrow G(U)$\ ,
satisfying the relation 
$r_{UV}\ f(U)=f(V)\ r_{UV}$\ .

	Morphisms of sheaves are 
simply morphisms of the underlying 
presheaves.

	Let $(X, {\cal{O}})$ be a 
complex manifold. A sheaf $B$ over $X$ is 
called a {\it coherent sheaf} 
of ${\cal{O}}$-modules if for each
$x \in X$ there is a neighborhood
$U$ of $x$ such that there is an exact 
sequence of sheaves over $U$,
\[0\rightarrow B|_{U}\rightarrow 
{\cal O}^{\oplus p_{1}}|_{U}
\rightarrow {\cal O}^{\oplus p_{2}}|_{U}
\rightarrow \ldots \rightarrow 
{\cal O}^{\oplus p_{k}}|_{U}
\rightarrow 0 \ .\]
\section{Complexes}
Let $B^{\bullet}$ denote
a {\it complex} of coherent
sheaves \cite{4.}
\[B^{\bullet}  :  \ \ldots 
\stackrel{d^{i-2}}\longrightarrow \ \ B^{i-1} \ \
\stackrel{d^{i-1}}\longrightarrow \ \ B^{i} \ \
\stackrel{d^{i}}\longrightarrow \ \ B^{i+1} \ \
\stackrel{d^{i+1}}\longrightarrow \ldots \ ,\]
where $d^{i}d^{i-1} = 0 $.

	{\it Cohomology groups} of the
complex $B^{\bullet}$ are defined as
\[H^{i}(B^{\bullet})={Ker\ d^{i}}/{Im\ d^{i-1}}\ 
.\]

	A morphism of complexes $f: B^{\bullet}
\rightarrow C^{\bullet}$ induces a 
morphism of cohomology groups 
$H(f): H^{\bullet}(B^{\bullet})
\rightarrow H^{\bullet}(C^{\bullet})$\ .

	If $H(f)$ is an isomorphism, the morphism
$f$ is said to be a {\it quasi-isomorphism}.

	If morphisms $f$ and $g$ are 
{\it homotopy equivalent} , then 
$H(f)=H(g)$ .
\section{Categories}
In this section we shall give some
formal definitions \cite{6.}.

	A {\it category} ${\cal {C}}$
consists of the following data\ :

	1)\ A class Ob ${\cal {C}}$
of objects $A$, $B$, $C$, $\ldots$\ ;

	2)\ A family of disjoint sets 
of morphisms Hom($A$, $B$), one for
each ordered pair $A$, $B$ of objects\ ;

	3)\ A family of maps
\[\mbox{Hom}(A,B)\times \mbox{Hom}(B, C)\rightarrow 
\mbox{Hom}(A, C)\ ,\]
one for each ordered triplet $A$, 
$B$, $C$ of objects.

	These data obey the axioms\ :

	a)\ If $f : A\rightarrow B\ ,\
g : B\rightarrow C\ ,\ h : C\rightarrow D\ ,$  
\ then composition of morphisms is associative, 
that is, $h(gf)=
(hg)f$\ ;

	b)\ To each object $B$ there 
exists a  morphism $1_{B}: B\rightarrow B$ 
such that $1_{B} f=f$\ , 
$g 1_{B}=g$\ \ for \ $f: 
A\rightarrow B$
\ and \ $g: B\rightarrow C$\ .\\

	An {\it additive category} is a
category in which each set of morphisms
Hom($A$, $B$) has the structure of an 
abelian group, subject to the
following axioms\ :

	{\bf A1}\ Composition of morphisms
is distributive, that is,
\[(g_{1}+g_{2})f = g_{1}f
+g_{2}f\ , \ \ \ h(g_{1}
+g_{2})=hg_{1}+ hg_{2}\]
for any $g_{1}, g_{2} : B
\rightarrow C\ , \ \ f : A\rightarrow
B\ ,\ \ h : C\rightarrow D$\ ;

	{\bf A2}\ There is a null 
object $0$ such that Hom($A$, $0$)
and Hom($0$,~$A$) consist of one
morphism for any $A$\ ;

	{\bf A3}\ To each pair of 
objects $A_{1}$ and $A_{2}$ there
exists an object $B$ and four morphisms
\[\hspace*{0mm}p_{1}\ \ \ \ \ \
p_{2}\]
\vspace*{-10mm}
\[ A_{1}\ \ {\cf}\ \  B\ \
{\cg}  \ \ A_{2}\]
\vspace*{-10mm}
\[\hspace*{-1mm}i_{1}\ \ \ \ \ \ \ i_{2}\]
\vspace*{-6mm}
which satisfy the identities\\
\[p_{1}i_{1}=1_{A_{1}}\ ,\ 
\ p_{2}i_{2}=1_{A_{2}}\ ,\ \
i_{1}p_{1}+i_{2}p_{2}=1_{B}\ ,\ 
\ p_{2}i_{1}=p_{1}i_{2}=0\ . \]

	An {\it abelian category}
${\cal {A}}$ is an additive category
which satisfies the additional
axiom\ :

	{\bf A4}\ To each morphism
$f : A\rightarrow B$ there
exists the sequence
\[K\ \stackrel{k}{\rightarrow}\  
A \ \stackrel{i}{\rightarrow}\
I\ \stackrel{j}{\rightarrow}\ B \ 
\stackrel{c}{\rightarrow}\ K^{'}\]
with the properties

	a)\ $ji=f$\ ,

	b)\ $K$ is a kernel
of $f$\ , $K^{'}$ is a
cokernel of $f$\ ,

	c)\ $I$ is a cokernel of
$k$ and a kernel of $c$\ .\\
The {\it category of
coherent sheaves} is the abelian
category ${\cal {A}}$\ .
\section{The derived category}
The derived category
D(${\cal {A}}$) is constructed
in three steps \cite{4.}\ :\\
{\it 1st step. } \hspace*{-3mm}The {\it
category of complexes} of coherent 
sheaves Kom(${\cal {A}}$) is
determined as follows\\
\hspace*{6mm}Ob Kom(${\cal {A}}$)\ =\ 
\{complexes $B^{\bullet}$
of coherent sheaves\}\ ,\\
\hspace*{6mm}Hom($B^{\bullet}$, $C^{\bullet}$) 
= \{morphisms of complexes
$B^{\bullet} \rightarrow 
C^{\bullet}$\}\ ;\\
{\it 2nd step.} The
{\it homotopy category} K(${\cal {A}}$)
is determined as follows\\
\hspace*{4.7mm}
Ob K(${\cal {A}}$) = Ob Kom(${\cal {A}}$)\ ,\\
\hspace*{6mm}Mor K(${\cal {A}}$) = 
Mor Kom(${\cal {A}}$) 
modulo homotopy equivalence\ ;\\
{\it 3rd step.} The
{\it derived category} D(${\cal {A}}$)
is determined as follows\\
\hspace*{4.7mm}
Ob D(${\cal {A}}$) = Ob 
K(${\cal {A}}$)\ ,\\
\hspace*{6mm}The morphisms of D(${\cal {A}}$)
are obtained from morphisms in\\
\hspace*{4.7mm} K(${\cal {A}}$) by inverting
all quasi-isomorphisms.\\
The  derived category 
D(${\cal {A}}$) is the additive
category.
\section{Triangulated structure}
The derived category
D(${\cal {A}}$) admits a
{\it triangulated structure} \cite{4.}
with {\it shift functor} $[n]$
defined by 
\[(B[n])^{i}=B^{n+i}\]
and with a class of 
{\it distinguished triangles}\\
\vspace*{-0.2cm}
\bec
\begin{tabular}{ccccc}
&&\hspace*{-6mm}
$C$&&\\
&&&&\vspace*{-10mm}\\
&$\hspace*{6mm}\stackrel{[1]}
{\swarrow}$&&$\hspace*{-6mm}
\nwarrow$&\hspace*{2cm} $C=A[1]\oplus B$\\
$A$\hspace*{-15mm}
&&\hspace*{-6mm}
$\longrightarrow$&&
\hspace*{-57mm}$B$
\end{tabular}
\ec
\vspace*{4mm}
These data satisfy a number
of axioms. The {\it octahedral
axiom} is an essential 
ingredient in the study of 
D-brane stability \cite{1.}. The
octahedral axiom states
that there exists the octahedron
consisting of a top cap and
a bottom cap\ :\\
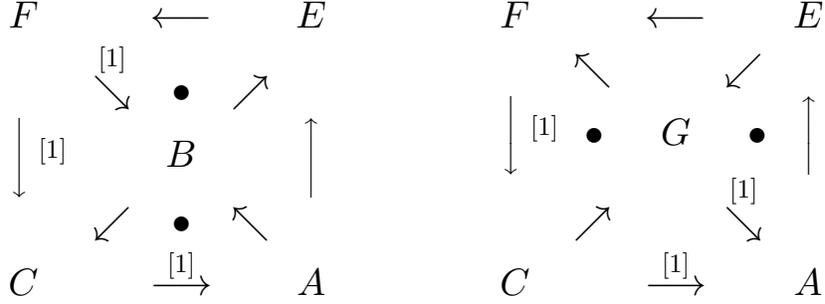
\[\begin{array}{ccc}
F&\longleftarrow&E\\
&\stackrel{[1]}{\searrow}
\hspace*{4mm}\bullet\hspace*{4mm}
\nearrow&\\
\hspace*{4mm}
\Bigg\downarrow
\stackrel{[1]}{}
&B&\Bigg\uparrow \\
&\swarrow\hspace*{4mm}
\bullet\hspace*{4mm}
\nwarrow & \\
C&\stackrel{[1]}
{\longrightarrow}&A\\
&&\\
\end{array}
\phantom{quality}\begin{array}{ccc}
F&\longleftarrow&E\\
&\hspace*{-2mm}\nwarrow
\hspace*{15mm}
\swarrow&\\
\hspace*{4mm}
\Bigg\downarrow
\stackrel{[1]}{}
&\bullet
\phantom{123}
G\phantom{123}
\bullet&
\Bigg\uparrow \\
&\hspace*{-2mm}
\nearrow\hspace*{15mm}
\stackrel{[1]}{\searrow}&  \\
C&\stackrel{[1]}
{\longrightarrow}&A\\
&&\\
\end{array} \]
\vspace*{-8mm}
\bec
(distinguished triangles are marked 
by $\bullet$)
\ec
\vspace*{4mm}
	
	Interpreting D-branes
as vertices of the octahedron,
it is possible to describe
{\it D-brane decays}\ :
if $C$ is stable against decay
into $A$ and $B$, but that $B$
itself is unstable with respect
to a decay into $E$ and $F$, 
than $C$ will always be unstable
with respect to decay into $F$
and some bound state $G$ of $A$ and $E$.
\section{The quintic}
Let $X$ be the quintic hypersurface
in $CP^{4}$ . The mirror $Y$ is defined as
the orbifold $X$/{\bf Z}$_{5}^{3}$ . 
In virtue of {\it mirror symmetry} 
\cite{7.} the {\it K$\ddot{a}$hler 
moduli space} of $X$ is identified with 
the {\it complex structure moduli
space} of $Y$. The complex structure
moduli space of $Y$ is described
by the Picard-Fuchs equation
\[\{\theta_{z}^{4}+5z(5\theta_{z}
+4)(5\theta_{z}+3)(5\theta_{z}
+2)(5\theta_{z}+1)\}
\omega_{k}(z)=0\ ,\]
where $\theta_{z}=z\ {d}/{dz}$ , 
the complex variable $z$
spans the complex structure
moduli space of $Y$.

	The {\it Landau-Ginzburg
point} of the moduli space of $X$
is mirror to $z=\infty$ , the
{\it large radius limit} of $X$
is mirror to $z=0$ , the {\it 
conifold point} of $X$ is mirror
to $z=1$. The periods $\omega_{k}(z)$
are singular at these three points.
\section{Monodromy}
Acting on the derived category
D(${\cal {A}}$), the {\it 
monodromy} is induced by a 
{\it Fourier-Mukai transform}
\cite{8.}
associated to some generator
$K^{\bullet}\in$ D(${\cal {A}}$).
The formula for the monodromy 
action on a complex $B^{\bullet}$
is
\[B^{\bullet}\longmapsto 
Rp_{1*}(K^{\bullet}
\stackrel{L}{\otimes}
p_{2}^{*}(B^{\bullet}))\ .\]
Geometry associated to this
monodromy action is
\bec
\begin{tabular}{ccccccc}
\vspace*{-1mm}
&&&$\bigtriangleup$&&& \\
&&&$\bigcap$&&& \\
\vspace*{-1mm}
&&&$X$$\times$$X$&&& \\
\vspace*{-1mm}
&&&\hspace*{-0.08cm}
$\stackrel{p_{1}}{\swarrow}$
\hspace*{1cm}
$\stackrel{p_{2}}{\searrow}$&&& \\
\vspace*{-1mm}
&&&$X$\hspace*{2cm}$X$&&& \\
\end{tabular} 
\ec
where $\bigtriangleup\subset 
X\times X$
is the diagonal embedding
of $X$ . 

	In the formula 
for the monodromy action, we

	1)\ Take a complex
of sheaves $B^{\bullet}$
on $X$, "pull it back"
to the inverse-image complex 
of sheaves $p_{2}^{*}(B^{\bullet})$
on $X\times X$\ ;

	2)\ Take the tensor-product
with the generator $K^{\bullet}$
and construct the left-derived
complex of sheaves\ ;

	3)\ "Push-forward"
to the direct image complex 
$p_{1\*}(\cdot)$ and construct
the right-derived complex of
sheaves on $X$\ . 

	The most obvious monodromy
is that about the Landau-Ginsburg 
point in the K$\ddot{a}$hler
moduli space of the quintic. This
monodromy is generated by \cite{1.}
\[K^{\bullet}_{LG}=\ 0\  
\rightarrow \
{\cal{O}}\ {\gf}\ {\cal{O}} (1)
 \ \rightarrow
\ {\cal{O}}_{\triangle} (1) \ 
\rightarrow \ 0 \ .\]
The monodromy calculations
for ${\cal {O}}\in$ D(${\cal {A}}$) 
yield the result\\
\[\hspace*{-4.7cm}M_{LG}({\cal{O}})
\hspace*{0.5cm} =\
 0 \rightarrow \ {\cal{O}}^{\oplus 5}
\ \rightarrow \ {\cal{O}} (1)\ 
\rightarrow \ 0\]
\[\hspace*{-2cm}(M_{LG})^{2}
({\cal{O}})= \ 0 \rightarrow
\ {\cal{O}}^{\oplus 10}\ \rightarrow
\ {\cal{O}}(1)^{\oplus 5}\ \rightarrow
\ {\cal{O}}(2)\ \rightarrow
\ 0 \]
\[(M_{LG})^{3}({\cal{O}})=\ 0 
\rightarrow
\ {\cal{O}}^{\oplus 10}\ \rightarrow
\ {\cal{O}}(1)^{\oplus 10}\rightarrow
\ {\cal{O}}(2)^{\oplus 5}\rightarrow 
\ {\cal{O}}(3)\
\rightarrow \ 0 \]
\[\hspace*{-8.4cm}(M_{LG})^{4}
({\cal{O}})=\ {\cal{O}}(-1)[4]\]
\[\hspace*{-9.4cm}(M_{LG})^{5}
({\cal{O}})=\ {\cal{O}}[2]\]
\section{Boundary linear
$\sigma$-model}
{\it Boundary linear $\sigma$-model}
\cite{2.}
is determined by the Lagrangian
\[\hspace*{-4.7cm}L=\sum\limits_{n}
\Biggl( i{\overline{\beta}}^{(2n)}
\partial_{0}\beta^{(2n)}+i{\overline
{\rho}}^{(2n+1)}
\partial_{0}\rho^{(2n+1)}\ +\]
\[\phantom{L} 
+\ \frac{1}{2}
{\overline{\beta}}^{(2n)}
(|\kappa^{(2n+1)}|^{2}
{\overline{c}}_{2n}
c_{2n}+|\kappa^{(2n)}|^{2}c_{2n-1}
{\overline{c}}_{2n-1})
\beta^{(2n)}\ +\]
\[\phantom{L=\sum\limits_{n}}
+\ \frac{1}{2}{\overline{\rho}}
^{(2n+1)}
(|\kappa^{(2n+2)}|^{2}{\overline{c}}
_{2n+1}c_{2n+1}
+|\kappa^{(2n+1)}|^{2}c_{2n}{\overline{c}}
_{2n})\rho^{(2n+1)}\Biggr)\ , \]
which involves superfields
${\beta}^{(2n)}$, $\rho^{(2n+1)}$,
$c_{k}$\ .

	Consider the complex of
direct sums of holomorphic line
bundles
\[\ldots\rightarrow
\bigoplus\limits_{i}{\cal{O}}
(m_{i}^{(2n-1)})
\stackrel{c_{2n-1}}{\longrightarrow}
\bigoplus\limits_{i}
{\cal{O}}(m_{i}^{(2n)})
\stackrel{c_{2n}}{\rightarrow}
\bigoplus\limits_{i}
{\cal{O}}(m_{i}^{(2n+1)})
\stackrel{c_{2n+1}}{\longrightarrow}
\ldots\]
Sections of holomorphic
line bundles describe superfields
${\beta}^{(2n)}$, $\rho^{(2n+1)}$\ ;
differentials describe superfields
$c_{k}$\ .\\
\vspace{1cm}\\
{\Large\bf Acknowledgement}\\

\hspace*{-6mm}This material 
was presented in the lecture
given by the author at the
Institute of Mathematics (Kiev, Ukraine).
The author thanks audience for
questions and comments.

\end{document}